\documentclass[twocolumn,aps,amsmath,amssymb,prl,epsf,showpacs,preprintnumbers]{revtex4}

\usepackage{psfig}
\usepackage{graphicx}
\usepackage{dcolumn}
\usepackage{bm}

\newcommand{\be}{\begin{equation}}
\newcommand{\ee}{\end{equation}}
\newcommand{\bq}{\begin{eqnarray}}
\newcommand{\eq}{\end{eqnarray}}

\usepackage{psfig}
\usepackage{graphicx}
\usepackage{dcolumn}
\usepackage{bm}

\begin{document}
\preprint{}
\title{Route of hydrodynamic turbulence in rotating shear flow: Secondary perturbation
triggers growing eigenmodes through the elliptical instability}

\author{Banibrata Mukhopadhyay\thanks{bmukhopa@cfa.harvard.edu}
}
\affiliation{Institute for Theory and Computation, Harvard-Smithsonian Center
for Astrophysics, 60 Garden Street, MS-51, Cambridge, MA 02138}

\begin{abstract}

Origin of hydrodynamic turbulence in rotating shear flow, e.g.
plane Couette flow including the Coriolis force, is a big puzzle.
While the flow often exhibits turbulence in laboratory
experiments, according to the linear perturbation theory it should always be
stable. We demonstrate that the secondary disturbance to the primarily
perturbed flow triggers
the elliptical instability in such a system and hence an
exponential growing eigenmode. This result has an immediate
application to astrophysics and geophysics which often exhibit a turbulent flow
with the Keplerian angular momentum profile. We address the origin of
turbulence in such a Keplerian flow which is similar to rotating
Couette flow.

\end{abstract}


\maketitle

Despite of many efforts devoted, the origin of hydrodynamic
turbulence is still poorly understood. One of the main problems
behind this is that there is a significant mismatch between the
predictions of linear theory and experimental data. For example,
in the case of plane Couette flow, laboratory experiments and
numerical simulations show that the flow may be turbulent for a
Reynolds number, $R\sim 350$, while according to the linear
theory the flow should be stable for all $R$. Similar mismatch
between theoretical results and observations is found in
geophysical and astrophysical contexts, where the accretion flow
of neutral gas with Keplerian angular momentum profile is a
common subject. A Keplerian accretion disk flow having a very low
molecular viscosity must generate turbulence and successively
diffusive viscosity, which support transfer of mass inward and
angular momentum outward. However, theoretically this
flow never exhibits any unstable mode which could trigger
turbulence in the system. On the other hand, laboratory
experiments of Taylor-Couette systems, which are similar to
Keplerian disks, seem to indicate that although the Coriolis force
delays the onset of turbulence, the flow is ultimately unstable
to turbulence for Reynolds numbers larger than a few thousand
\cite{richard2001}, even for subcritical systems. Longaretti
\cite{long} reviews the experimental evidence for the existence
of turbulence in subcritical laboratory systems, and concludes,
based on phenomenological analogy, that a similar process must
happen in astrophysical accretion flows. Indeed, Bech \& Anderson
\cite{bech} see turbulence persisting in numerical simulations of
subcritical rotating flows for large enough Reynolds numbers.

How does shearing flow that is linearly stable to perturbations
switch to a turbulent state?  Since last decade, many authors
have come forward with a possible explanation of this fact, based
on {\it bypass} transition (see
\cite{butfar,rh,tref,chagelish,umurhan,man} and references
therein), where the decaying linear modes show an arbitrarily
large transient energy growth at a suitably tuned perturbation.
In lieu of linear instabilities e.g. magnetorotational
instability, the transient energy growth, supplemented by a
non-linear feedback process to repopulate the growing
disturbance, could plausibly sustain turbulence for large enough
Reynolds numbers.

However, in the case of rotating shear flow, e.g. Couette-Taylor
flow, Keplerian accretion flow, the transient energy growth is
insignificant for three-dimensional perturbations. The Coriolis
effect, which absorbs the pressure fluctuation, is the main
culprit to kill any growth of energy. Nevertheless, at a very
large Reynolds number, certain two-dimensional perturbations may
produce large transient growth in such a flow. However, in 
two-dimension, the underlying perturbations must ultimately decline 
to zero in the presence of viscosity \cite{umurhan,man}.
To overcome this limitation, it is necessary to invoke three-dimensional effects.
Various kinds of secondary instabilities, such as the elliptical
instability, are widely discussed as a possible route to
self-sustained turbulence in linearly perturbed shear flows (see,
e.g. \cite{pier86,bay86,cc86,ls87,helor88,wal90,
c89,lrm96,ker02}). These effects, which generate
three-dimensional instabilities of a two-dimensional flow with
elliptical streamlines, have been proposed as a generic mechanism
for the breakdown of many two-dimensional high Reynolds number
flows whose vortex structures can be locally seen as elliptical
streamlines. However, to our knowledge, such effects have not been discussed properly
in literatures for rotating Keplerian flows which
have a vast application to various natural phenomena.
More than a decade ago, the possible role of the
elliptical instability in astrophysical accretion disk physics
was first explored \cite{good} and it was shown that angular
momentum may be transferred from the disk to the tidal source by
the instability effect \cite{lub,rygood}. Therefore, naturally we
are motivated to see whether these three-dimensional instabilities
are present in rotating shear flows which consist of elliptical
streamlines under two-dimensional perturbation. 
We plan to show that in presence of secondary effects, three-dimensional 
perturbation can generate large growth and presumably trigger turbulence
in rotating shear flows. Unlike the growth of transient kind which is significant 
in two-dimension only, in the present case it is essentially the three-dimensional 
effects which produce large growth. Possibility of large growth in shear flows with
{\it rotation} by three-dimensional perturbation opens a new window to explain
hydrodynamic turbulence.

Let us consider a small patch of rotating shear flow whose unperturbed
velocity profile corresponds to a linear shear of the form $\vec{U}_0=(0,-x,0)$.
Because of rotation, a Coriolis acceleration acts on the fluid element
having angular frequency $\vec{\Omega}=(0,0,\Omega_3)$ where according to choice
of our units $\Omega_3=1/q$. The parameter $q$ is positive, describing the
radial dependence of $\Omega(r)=\Omega_0(r_0/r)^q$ (see \cite{man} for detail).
We choose all the variables to be dimensionless. When
the Reynolds number is very large, this flow exhibits a large transient growth under
a suitable two-dimensional perturbation \cite{man} which modifies the 
linear shear profile as
\begin{eqnarray}
\vec{U}&=&\vec{U}_0+\vec{w}=(w_x,-x+w_y,0)={\bf A}.\vec{d},\\
\nonumber
w_x&=&\zeta\frac{k_y}{\kappa^2}\sin(k_xx+k_yy),\,\,w_y=-
\zeta\frac{k_x}{\kappa^2}\sin(k_xx+k_yy),
\label{primper}
\end{eqnarray}
where $\vec{d}$ is the position vector and $\bf A$ is a tensor of rank $2$. 
Now expanding the $\sin(k_xx+k_yy)$ terms and choosing a small patch, 
$\bf A$ is given by
\begin{equation}
{\bf A}=A_j^k=\left(\begin{array}{ccc} \zeta\sqrt{\epsilon(1-\epsilon)} &
\zeta(1-\epsilon) & 0\\ -(1+\zeta\epsilon) &
-\zeta\sqrt{\epsilon(1-\epsilon)} & 0\\ 0 & 0 & 0 \end{array}\right),
\label{axy}
\end{equation}
where $\epsilon=(k_x/\kappa)^2$, $\kappa=\sqrt{k_x^2+k_y^2}$, and
$\zeta$ is the amplitude of vorticity perturbation. Here
$k_x=k_{x0}+k_yt$, which basically is the radial component of
wave-vector, varies from $-\infty$ to $+\infty$, where $k_{x0}$
is a large negative number. The above plane wave typed perturbation
is frozen into the fluid and is sheared along with the
background flow. At $t=0$, the effective wave vector of
perturbation in the $x$ direction ($k_x$) is negative, which
provides very asymmetric leading waves.  As time goes on, the
wavefronts are straightened out by the shear and $|k_x|$
decreases. At the time when the transient growth is maximum,
$k_x\sim 0$ and the wavefronts become almost radial. At yet later
time, the growth decreases and the wave becomes of a trailing
pattern. Clearly $\vec{U}$ describes a flow having generalized
elliptical streamlines with $\epsilon$, a parameter related to the measure of
eccentricity \footnote{Note that $\epsilon$ is a parameter related to the measure
of eccentricity but not the eccentricity itself.},
runs from $0$ to $1$ as the perturbation evolves.
Now we plan to study how does this perturbed shear flow behave
under the further perturbation, namely a {\it secondary
perturbation}.

The linearized equations for the evolution of a secondary perturbation
$\vec{u}$, such that $\vec{U}\rightarrow \vec{U}+\vec{u}$, to the flow are
\begin{eqnarray}
\nonumber
&&\left(\partial_t+\vec{U}.\nabla\right)\vec{u}+\vec{u}.\nabla\vec{U}+2\vec{\Omega}\times\vec{u}
=-\nabla\tilde{p}+\frac{1}{R}\nabla^2\vec{u},\\
&&\nabla.\vec{u}=0
\label{secbase}
\end{eqnarray}
where $\tilde{p}$ is pressure in the fluid element and $R$ is the Reynolds number.
A secondary perturbation of the plane wave kind is given by
\begin{equation}
(u_i,\tilde{p})=(v_i(t),p(t)) \exp(ik_m(t) x^m),
\label{pet}
\end{equation}
where the Latin indices, run from $1$ to $3$, indicate the spatial components of
a variable, e.g. $x^m\equiv (x,y,z)$. Therefore, after some algebra,
we obtain the evolution of a linearized secondary perturbation
\begin{equation}
\dot{v}_j+A_j^k\,v_k+2\,\epsilon_{mkj}\Omega^m v^k=-ip\,k_j
-\frac{v_j}{R}\,k^2,
\label{perteq}
\end{equation}
\begin{equation}
\dot{k}_j=-(A^m_j)^T\,k_m,\,\,\,\,\,
k^n\dot{v}_n=k^m\,A^n_m\,v_n,
\label{keq}
\end{equation}
where `over-dot' indicates a derivative with respect to $t$,
$\epsilon_{mkj}$ is a Levi-Civita tensor, and $k^2=k_mk^m$.
A similar set of equations was obtained by Bayly \cite{bay86}, except that they now have additional
terms induced due to the Coriolis and the viscous effects along with a modified $\bf A$.
For a fixed $k_x$, the components $k_1$ and $k_2$ of wave-vector [$\vec{k}=k_m=(k_1,k_2,k_3)$] of
secondary perturbation oscillate in time with the angular frequency
$\varpi=\sqrt{\zeta(1-\epsilon)}$, while the vertical component, $k_3$,
is constant.
As we choose the signature of the background Minkowski space-time to be $[-,+,+,+]$,
it does not matter whether
any Latin index appears as a lower case or an upper case. For example,
$A^k_j=A_{jk}$, where $j$ and $k$ indicate the row number and the column
number of the associated matrix respectively.
Projecting out eqn. (\ref{perteq}) by $P^j_i=\delta^j_i-k^{-2}k^j\,k_i$ and
using eqn. (\ref{keq}), we obtain
\begin{eqnarray}
\nonumber
\dot{v}_i=\left(2\frac{k^j\,k_i}{k^2}-\delta^j_i\right)A^k_j\,v_k
-2\,\epsilon_{mki}\,\Omega^m\,v^k-\frac{v_i}{R}k^2\\
+\left(2\,\epsilon_{mkj}\,\Omega^m\,v^k+\frac{v_j}{R}k^2\right)\frac{k^jk_i}{k^2}.
\label{veq}
\end{eqnarray}
Now we specifically concentrate on the flow having low viscosity. Therefore,
we neglect the viscous term in eqn. (\ref{veq}) comparing others and
rewrite
\begin{widetext}
\begin{equation}
\small
\dot{v}_i=\Lambda^j_i\,v_j,
\label{vmat}
\hskip0.3cm \Lambda_{ji}=\left(\begin{array}{ccc} \left(\frac{2k_1^2}{k^2}-1\right)A_{11}+
\frac{2k_1k_2}{k^2}\left(A_{21}+\Omega_3\right) & \frac{2k_1^2}{k^2}\left(A_{12}-\Omega_3
\right)+2\Omega_3-A_{12}+\frac{2k_1k_2}{k^2}A_{22}
 & 0\\ \frac{2k_1k_2}{k^2}A_{11}+\frac{2k_2^2}{k^2}\left(A_{21}+\Omega_3\right)
-A_{21}-2\Omega_3 & \frac{2k_1k_2}{k^2}\left(A_{12}-\Omega_3\right)+
\left(\frac{2k_2^2}{k^2}-1\right)A_{22} &
0\\ \frac{2k_1k_3}{k^2}A_{11}+\frac{2k_2k_3}{k^2}\left(A_{21}+\Omega_3\right) &
\frac{2k_1k_3}{k^2}\left(A_{12}-\Omega_3\right)+\frac{2k_2k_3}{k^2}A_{22}
& 0 \end{array}\right).
\end{equation}
\end{widetext}
The general solution of eqn. (\ref{vmat}) for a particular $k_x$
can be written as a linear superposition of Floquet modes
\begin{equation}
v_i(t)=\exp(\sigma\,t)\,f_i(\phi),
\label{flo}
\end{equation}
where $\phi=\varpi\,t$, $f_i(\phi)$ is a periodic function having
time-period $T=2\pi/\varpi$, and $\sigma$ is the Floquet exponent
which is the eigenvalue of the problem. As $k_x$ varies, $\epsilon$
changes, therefore the eigenvalue changes. Clearly, if $\sigma$ is positive then
the system is unstable. To compute $\sigma$, one has to evaluate the elements
of a matrix $M_{mi}(2\pi)$ whose eigenvalues and eigenvectors are respectively
$e^{\sigma T}$ and $f_i(2\pi)$, while $M_{mi}(\phi)$ itself satisfies
\begin{equation}
\frac{dM_{ji}(t)}{dt}=\Lambda^m_jM_{mi}(t),
\label{matevo}
\end{equation}
with the condition $M_{ji}(0)=\delta_{ji}$. Clearly, (\ref{matevo}) is the 
evolution equation of velocity. If $\mu$ is a real positive
eigenvalue of $M_{mi}(2\pi)$, then the corresponding
$\sigma=ln(\mu)/T$. One can evaluate $\mu$ by an elementary numerical technique. However,
eqn. (\ref{matevo}) has exact analytical solution for an
initial perturbation $\vec{k}_0=(0,0,1)$. In this case, $\vec{k}$
remains constant throughout for a fixed $\epsilon$ and therefore the growth rate,
$\sigma$, is the highest eigenvalue of $\Lambda_j^m$ given by
\begin{equation}
\sigma=\sqrt{\zeta\epsilon-(2\Omega_3-1)(2\Omega_3-\zeta)}.
\label{sigcon}
\end{equation}

$\bullet$ When $\Omega_3=0$, $\sigma=\sqrt{\zeta(\epsilon-1)}$. This verifies
that non-rotating two-dimensional shear flow is always hydrodynamically
stable under a {\it pure} vertical perturbation.

$\bullet$ When $\Omega_3=1/2$, $\sigma=\sqrt{\zeta\epsilon}$. Therefore, rotating
shear flow with a constant angular momentum profile is always hydrodynamically unstable.
The energy growth rate of perturbation increases with the strain rate, i.e.
eccentricity, of the flow.

$\bullet$ When $\Omega_3=2/3$, $\sigma=\sqrt{\zeta\epsilon-(4-3\zeta)/9}$.
Therefore, a Keplerian flow with elliptical streamlines 
is hydrodynamically unstable under a {\it pure} vertical perturbation, only if $\zeta>1/3$.

\begin{figure}
\begin{center}
\vskip-2.0cm
\includegraphics[width=10cm,height=8cm]{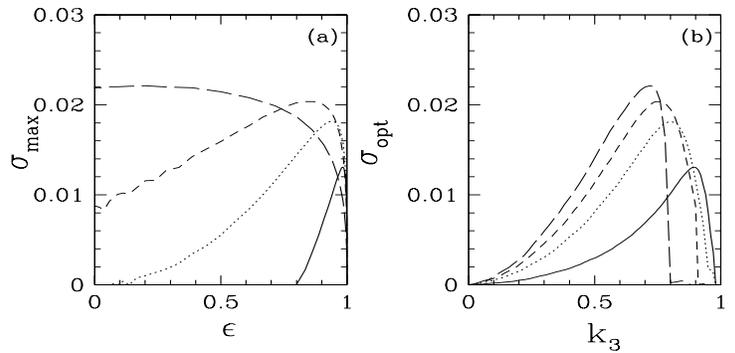}
\end{center}
\vskip-1.5cm \caption{(a) Variation of maximum growth rate as a
function of eccentricity parameter when solid, dotted, dashed and
long-dashed curves indicate the results for $\zeta=0.2,0.1,
0.05,0.01$ respectively. 
(b) Variation of optimum growth rate
as a function of vertical component of perturbation wave-vector
when various curves are same as of (a). $q=3/2$ throughout.
} \label{fig1}
\end{figure}

However, apart from the vertical one, 
there are some other three-dimensional perturbations \footnote{By vertical perturbation
we mean that only the vertical component of initial perturbation wave-vector is non-zero,
while any perturbation with a non-zero vertical component of initial wave-vector
is called three-dimensional perturbation.}
which can generate instability
in rotating shear flows with $\zeta < 1/3$, that we describe by
numerical solutions. We essentially focus on the Keplerian flow when $q=3/2$.
As primary perturbation evolves with time,
eccentricity decreases, and then energy growth rate due to
secondary perturbation changes. Figure \ref{fig1}a shows the
variation of maximum growth rate, $\sigma_{max}$, as a function of
eccentricity parameter, $\epsilon$, for various values of $\zeta$, the
amplitude of primary perturbation, in the Keplerian flows. By ``maximum'' we refer the
quantity obtained by maximizing over the vertical component of
wave-vector, $k_{3}$.
While a vertical perturbation gives rise to the best growth rate
for $\zeta>1/3$, as we show above analytically, for $\zeta < 1/3$ other
three-dimensional perturbations maximize growth rate. At small $\epsilon$
and large $\zeta$, the streamlines of the flow essentially become circular 
(see eqn. (\ref{axy})), and thus in absence of any significant elliptical effect
growth rate severely decreases. On the other hand, when $\epsilon$
and $\zeta$ both are small, the background reduces to plane shear
structure and therefore any growth arises due to primary perturbation only.

Figure \ref{fig1}b shows the variation of optimum growth rate,
$\sigma_{opt}$, as a function of $k_3$ for various values of $\zeta$, in the
Keplerian flows.
By ``optimum'' we refer the quantity obtained by maximizing over
$\epsilon$. Interesting fact to note is that the optimum growth rate is
obtained always for three-dimensional perturbation with
significant vertical effect, i.e. a non-zero value of $k_3$.
Moreover, as $\zeta$ increases, instability occurs at larger $\epsilon$ with 
higher $k_3$. Therefore, three-dimensional elliptical instability is more prompt
at larger $\zeta$. 

Above results verify that at a range of $\epsilon$ the
{\it instantaneous} three-dimensional growth rate due to
secondary perturbation is always real and positive, which motivates us to
compute the actual growth itself during simultaneous evolution of
both the perturbations. As primary perturbation evolves, $k_x$
varies with time and therefore $\bf A$ does so. Thus, in reality,
the wave-vector of secondary disturbance is no longer periodic,
though the corresponding equation (\ref{keq}) is still valid. To
compute growth in energy one has to find out the elements of a
matrix
\begin{equation}
{\cal M}_{ik}(t)=M_{im}(t)^T M_{mk}(t)
\label{m2}
\end{equation}
whose largest eigenvalue is growth in energy at a time $t$. Clearly,
${\cal M}_{ik}(t)$ is the instantaneous energy of the flow.
$k_x$ and $k_1$ start with a large negative value when the flow is
highly eccentric. With time $k_x$ (as well as $k_1$) decreases in
magnitude and finally becomes zero when the streamlines become
circular. If $\zeta=0$, then $k_x$ is same as $k_1$. However, for
$\zeta>0$, $k_1$ increases faster than $k_x$, as follows from 
eqn. (\ref{keq}), and becomes positive during the evolution of
perturbation. When $k_1\rightarrow 0$, growth maximizes.

\begin{figure}
\begin{center}
\vskip-2.0cm
\includegraphics[width=10cm,height=8cm]{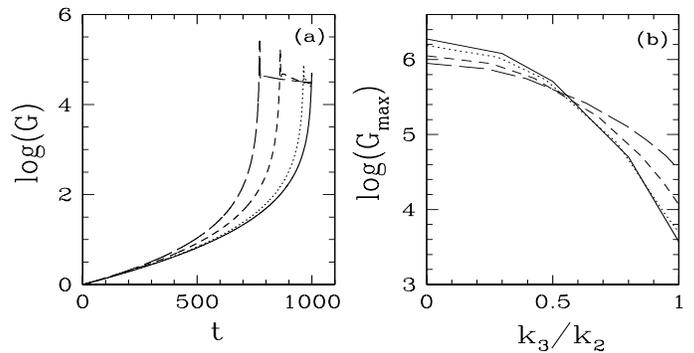}
\end{center}
\vskip-1.5cm \caption{(a) Variation of growth as a
function of time due to simultaneous evolution of both perturbations, 
when $k_3=0.8$. Solid, dotted, dashed and
long-dashed curves indicate the results for $\zeta=0,0.01,
0.05,0.1$ respectively. 
(b) Variation of maximum growth 
as a function of $k_3/k_2$, 
when various curves are same as of (a). Other parameters are 
$k_{x0}=k_{10}=-1000$, $k_{20}=1$, and $q=3/2$.
} \label{fig2}
\end{figure}

Figure \ref{fig2}a shows the variation of growth as a function of $t$ for
various values of $\zeta$. Note that as $\zeta$ increases, growth
maximizes at an earlier time. Figure \ref{fig2}b shows the variation of maximum
growth as a function $k_3/k_2$. The quantity $k_3/k_2$ carries
information about how three-dimensional the flow is. We know
that in two-dimensional plane shear flow the maximum
growth, $G_{max}$, scales with $k_{x0}^2$ \cite{man}.
However, for $\zeta>0$, $G_{max}$
decreases at small $k_3/k_2$, while increases at large $k_3/k_2$.
This clearly proves that three-dimensional secondary perturbation
triggers elliptical instability which produces larger growth. 
As $k_{x0}$ and $k_{10}$ ($\sim R^{1/3}$ \cite{umurhan,man})
increase, the effects due to elliptical instability increase,
and thus the corresponding growth does so. When $k_{x0}=k_{10}=-10^3$,
$G_{max}\sim 4\times 10^4$ at $k_3/k_2=1$ for $\zeta=0.1$, which is an order of 
magnitude larger compared to that for $\zeta=0$. If we consider a smaller $R$ with
the corresponding $k_{x0}=k_{10}=-10^2$, then $G_{max}$ at $k_3/k_2=1$ decreases
to $\sim 2\times 10^3$ for $\zeta=0.1$, which is still a factor of two larger 
compared to that for $\zeta=0$. Therefore, this is confirmed without any 
doubt that three-dimensional elliptical instability efficiently triggers turbulence 
in shear flows.

When $t$ increases from $0$ to $t_{max}=-k_{x0}/k_y$, $k_x(t)$
varies from $k_{x0}$ ($\sim -\infty$) to $0$, and $\epsilon$
decreases from $1$ to $0$. Note that 
$G_{max}\sim 2\times 10^3-4\times 10^4$ when $k_{10}^3\sim R\sim 10^6-10^9$.
As this large growth is the result of three-dimensional perturbation, even in presence
of viscosity the underlying perturbation effect should survive.
Presumably, this
growth factor is enough to trigger non-linear effects and turbulence in such flows.
There are many important natural phenomena where the Reynolds number is very large.
In astrophysical accretion disks $R$ always could be $10^{10}-10^{14}$ \cite{man}
because of their very low molecular viscosity. Therefore, the
present mechanism can have a very good application to 
such disk flows to explain their {\it turbulence puzzle} when it is cold
and neutral in charge. On the other hand, we argue that the subcritical transition
to turbulence in Couette flow may be the result of secondary perturbation which
triggers elliptical instability modes in the system. It is to be seen now 
whether all shear flows, exhibit subcritical turbulence in the laboratory,   
do also exhibit large growth due to secondary perturbation, according to the present 
mechanism.


The author is grateful to Ramesh Narayan for suggesting this problem and for
extensive discussion and encouragement
throughout the course of the work.
This work was supported in part by NASA grant NNG04GL38G and NSF grant
AST 0307433.

\end{document}